\documentclass[%
 reprint,
 superscriptaddress,
 amsmath,amssymb,
 aps,
 floatfix,
]{revtex4-2}
\usepackage{graphicx}
\usepackage{dcolumn}
\usepackage{bm}
\usepackage{hyperref}
\usepackage{here}
\usepackage{booktabs}
\usepackage{siunitx}
\usepackage{comment}
\usepackage[version=4]{mhchem}
\usepackage{amsmath,amssymb}
\usepackage{array}
\newcolumntype{H}{>{\setbox0=\hbox\bgroup}l<{\egroup}@{}}
\usepackage{xtab,afterpage}
\usepackage{tabularx}
\usepackage{lipsum}
\usepackage{longtable}
\usepackage{ragged2e}
\usepackage{placeins}
\usepackage{subfiles}
\newcolumntype{Y}{>{\RaggedRight\arraybackslash}X}
\newcolumntype{P}[1]{>{\RaggedRight\arraybackslash}p{#1}}

\begin{document}

\preprint{APS/123-QED}

\title{Expanding Universal Machine Learning Interatomic Potentials to 97 Elements Towards Nuclear Applications}

\author{Naoya Kuroda}
\thanks{These two authors contributed equally to this work.}
\affiliation{Center for Quantum Information and Quantum Biology, The University of Osaka, 1-2 Machikaneyama, Toyonaka 560-8531, Japan}
\author{Kenji Ishihara}
\thanks{These two authors contributed equally to this work.}
\affiliation{Center for Quantum Information and Quantum Biology, The University of Osaka, 1-2 Machikaneyama, Toyonaka 560-8531, Japan}
\affiliation{Graduate School of Science, The University of Osaka,
1-1 Machikaneyama, Toyonaka, Osaka 560-
0043, Japan}
\author{Tomoya Shiota}
\email{shiota.tomoya.ss@gmail.com}
\affiliation{Center for Quantum Information and Quantum Biology, The University of Osaka, 1-2 Machikaneyama, Toyonaka 560-8531, Japan} 
\author{Wataru Mizukami}
\email{mizukami.wataru.qiqb@osaka-u.ac.jp}
\affiliation{Center for Quantum Information and Quantum Biology, The University of Osaka, 1-2 Machikaneyama, Toyonaka 560-8531, Japan} 

\date{\today}

\begin{abstract}

Machine learning interatomic potentials (MLIPs) evaluate potential energy surfaces orders of magnitude faster while maintaining accuracy comparable to first-principles calculations, and universal MLIPs that cover most of the periodic table are becoming increasingly commonplace. However, existing large-scale datasets have limited or no coverage of heavy elements such as minor actinides crucial in the nuclear field, and universal MLIPs are typically limited to 89 elements. Here, we constructed a heavy element dataset HE26 containing minor actinides, based on experimental and computational literature data. By integrating this with existing molecular and crystal datasets, we developed an open-source universal MLIP covering 97 elements, the broadest elemental coverage to date. The resulting model showed strong performance on the inorganic MPtrj and organic OFF23 test sets and promising accuracy on HE26. The dataset and model open a pathway toward the development of energy resources and the design of novel materials, such as actinide-based high-entropy ceramics, in the nuclear field.

\end{abstract}

\maketitle

\section{\label{sec:intro}Introduction}

\begin{figure*}[t]
\includegraphics[width=2\columnwidth]{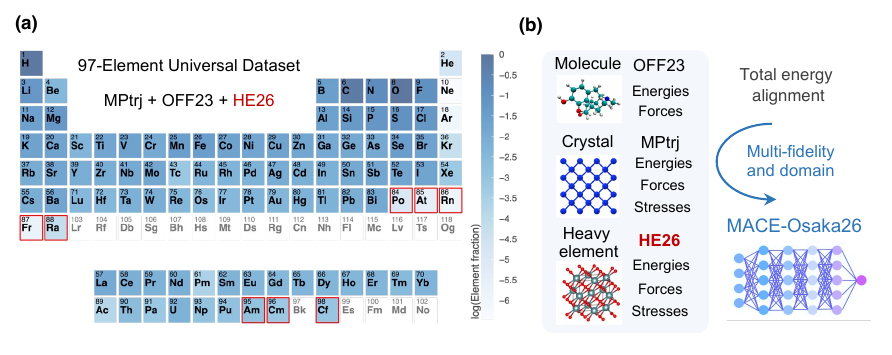}
\caption{\label{fig:1} Overview of the 97-element universal dataset and the MACE-Osaka26 model. (a) Periodic table showing the element distribution (log scale) of the combined dataset consisting of MPtrj, OFF23, and our HE26 dataset. Red boxes indicate the heavy elements newly added in this study. (b) Schematic of the training pipeline integrating molecular (OFF23) and crystal (MPtrj, HE26) data via the total energy alignment protocol.}
\end{figure*}

Minor actinide elements, including transuranium elements such as Am, Cm, and Cf, are not merely nuclear waste but key resources for the sustainability of the nuclear fuel cycle \cite{crawford2007_sfr_fuels_us_perspective, magni2021_minor_actinide_mox_tc_melting, oecdnea2012_hvh_recycling_tru_fast_reactors, kooyman2021_pt_review, oecdnea2011_benefits_impacts_actinide_pt}, and are emerging as crucial components in advanced applications ranging from americium oxide for space exploration radioisotope thermoelectric generators \cite{ambrosi2019_european_rtg_rhu_am241, brown2018_amppex_space_power, watkinson2021_am_oxide_rps_thermo} to actinide-bearing high-entropy ceramics designed for extreme environments \cite{Ward2024HECeramics, yang2021_u_he_pyrochlore, liu2024_u_doped_he_pyrochlore}.
Unfortunately, due to their strong radioactivity and toxicity, experimental measurements of important thermophysical properties, such as lattice thermal conductivity, linear thermal expansion, and viscosity, are extremely difficult and expensive, leaving the thermophysical and thermochemical properties of many minor actinide compounds largely unexplored \cite{minato2009thermochemical, njifon2020_mox_phonons_thermophysical, bonatti2025smaller, CARBAJO2001181, Konings2006, UCHIDA2009229, SOBOLEV200945, hurley2022_thermal_transport_review,WOODLEY1974103,SMITH2022119426,tkacheva2023viscosity, doi:10.1021/acs.jpcb.5c04764}.
Theoretical and computational approaches are therefore essential for characterizing these materials.

Computational prediction of thermophysical properties relies on the accuracy of the underlying interatomic potential.
First-principles calculations based on density functional theory (DFT) can model interatomic forces accurately, but their computational cost becomes prohibitive for multi-component systems: the reduced crystal symmetry requires large supercells, and the number of atomic displacement configurations needed for higher-order anharmonic force constants grows rapidly \cite{Togo2023JPSJ}.
Machine learning interatomic potentials (MLIPs) offer a practical route to overcome this bottleneck \cite{Behler2007PRL, Bartok2010PRL, Zhang2018PRL_DeePMD, jacobs2025practical, Shapeev2016MTP, drautz2019ace}.
MLIPs evaluate potential-energy surfaces orders of magnitude faster than DFT while maintaining near-DFT accuracy.

System-specific MLIPs, trained on first-principles data for particular systems, have recently predicted the high-temperature thermophysical properties of nuclear materials such as \ce{UO2} and \ce{ThO2} with accuracy approaching that of DFT \cite{stippell2024_uo2_dftu_mlip, dubois2024_uo2_mlip, kobayashi2022_tho2_mlip_scirep, AlzateVargas2025MLST, zhong2026_uo2_nep_tc, minamitani2019simulating}.
However, because these models are trained on individual systems, they are not suitable for screening across broad compositional spaces, such as actinide-containing mixed oxide fuels \cite{crawford2007_sfr_fuels_us_perspective, magni2021_minor_actinide_mox_tc_melting, oecdnea2012_hvh_recycling_tru_fast_reactors, kooyman2021_pt_review, oecdnea2011_benefits_impacts_actinide_pt}.

Universal MLIPs, also known as atomistic foundation models, are emerging to address this limitation as transferable interatomic potentials that cover a wide range of elements \cite{jacobs2025practical}.
Architectural developments, such as graph neural networks (GNNs)~\cite{schutt2018schnet} and $E(3)$-equivariant message passing~\cite{batzner20223, Musaelian2023Allegro, batatia2022mace}, together with advances in distributed scalability~\cite{park2024scalable} and more efficient equivariant architectures~\cite{D5SC07248D}, have driven recent progress.
Pretraining with large-scale open datasets~\cite{deng2023chgnet,BarrosoLuque2024OMat24,schmidt2022dataset, schmidt2024improving,Mazitov2025MAD} and heterogeneous training datasets~\cite{Shiota2025MACEOsaka24TEA, wood2025umafamilyuniversalmodels,batatia2025crosslearningelectronicstructure,takamoto2022towards,nomura2025allegrofmequivariantfoundationmodel,kim2025optimizingcrossdomaintransferuniversal,liang2025nep89universalneuroevolutionpotential} has further improved these models, enabling application to molecules, crystals, and surface systems.
However, existing universal MLIPs typically do not cover transuranium elements such as Am, Cm, and Cf \cite{Shiota2025MACEOsaka24TEA,nomura2025allegrofmequivariantfoundationmodel,takamoto2022towards,wood2025umafamilyuniversalmodels,batatia2024foundationmodelatomisticmaterials,batatia2025crosslearningelectronicstructure,kim2025optimizingcrossdomaintransferuniversal,liang2025nep89universalneuroevolutionpotential}.
This gap comes from the limited or absent coverage of heavy elements in current first-principles datasets \cite{jain2013commentary,deng2023chgnet}, which restricts the applicability of pretrained universal models not only to actinide mixed oxides but even to many heavy-element compounds.

To expand the elemental coverage of a universal MLIP, we develop a DFT dataset, Heavy Element 2026 (HE26). HE26 spans 65 elements in total and contains eight heavy elements (Am, Cm, Cf, Fr, At, Ra, Po, and Rn) that are rarely present in existing large-scale datasets.
The HE26 dataset consists of three subsets: (1) Basic Heavy Element (BHE): elemental solids and binary oxides; (2) Complex Heavy Element (CHE): diverse multicomponent systems; and (3) Compositionally Complex Fluorite Oxides (CCFO): binary-to-quinary fluorite-type mixed oxides of actinides, lanthanides, and rare earth elements, targeting actinide-based high-entropy ceramics and minor-actinide mixed-oxide fuels.
We integrate HE26 with the OFF23~\cite{kovacs2023mace} and MPtrj~\cite{batatia2024foundationmodelatomisticmaterials} datasets unified via the total-energy alignment (TEA) protocol~\cite{Shiota2025MACEOsaka24TEA}. Using this combined dataset, we train MACE-Osaka26, an open-source universal MLIP covering 97 elements.
We demonstrate the model's applicability by predicting lattice parameters and thermal conductivity of actinide fluorite oxides, including mixed oxides containing minor actinides.

The remainder of this paper is organized as follows. Section~\ref{sec:results} presents results in three parts: cross-domain accuracy of MACE-Osaka26 across crystalline, molecular, and heavy element datasets; detailed evaluation on the HE26 dataset; and lattice thermal conductivity predictions for actinide fluorite oxides. Section~\ref{sec:discussion} discusses the origins of the model's performance and remaining limitations. Section~\ref{sec:conclusion} summarizes our findings and outlook. Section~\ref{sec:methods} describes the dataset construction, MLIP training, and thermal conductivity calculation procedures. 

\section{\label{sec:results}Results}

\begin{figure*}[h]
\includegraphics[width=2\columnwidth]{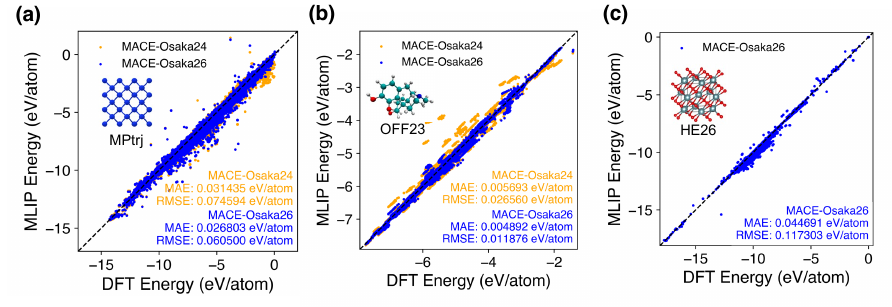}
\caption{\label{fig:2} Accuracy comparison between MACE-Osaka24 and MACE-Osaka26 on cross-domain systems. (a) Evaluation on the MPtrj training set representing crystalline systems, and (b) the OFF23 training set representing molecular systems, using both MACE-Osaka24 (orange) and MACE-Osaka26 (blue) models. (c) Results for the HE26 training set containing heavy elements. Note that only MACE-Osaka26 is shown in (c) as it incorporates eight heavy elements not supported by the MACE-Osaka24 model. Mean absolute errors (MAE) and root-mean-square errors (RMSE) are provided in eV/atom. Here, the main difference in the architecture between MACE-Osaka24 and MACE-Osaka26 is that their graph cutoff values are 4.5 Å and 6.0 Å, respectively.}
\end{figure*}

We first evaluate the cross-domain generalization capability of the MACE-Osaka26 model across molecules, crystals, and heavy-element compounds spanning 97 elements.
Table~\ref{tab:1} and Figure~\ref{fig:2} compare the prediction accuracies of MACE-Osaka24 and MACE-Osaka26 across crystalline (MPtrj), molecular (OFF23), and heavy-element (HE26) datasets. 
Appendix~\ref{sec:appendix:B} provides detailed descriptions of the prediction accuracies for energies on the test sets, and for forces and stresses on both the training and test sets.
As shown in Table~\ref{tab:1}, the prediction accuracies for the training and test sets exhibit generally similar trends across the MPtrj and OFF23 datasets. Furthermore, due to the limited total number of structures in the HE26 dataset, a separate test set was not partitioned. For these reasons, the parity plots illustrating the energy prediction accuracies in Figure~\ref{fig:2} are presented using the training sets for all three datasets.

For the crystalline MPtrj dataset, MACE-Osaka26 shows consistent improvements in energy prediction on the test set, with the mean absolute error (MAE) decreasing from 33.7 to 29.3 meV/atom and the root-mean-square error (RMSE) from 77.2 to 62.0 meV/atom. While the force MAE slightly increased, the force RMSE and stress RMSE showed meaningful reductions, indicating better handling of outlier configurations in bulk systems. The most striking improvement is observed in the molecular OFF23 test set. 
MACE-Osaka26 significantly enhanced energy accuracy compared to MACE-Osaka24, notably halving the RMSE from 25.2 to 12.3 meV/atom. 
This significant improvement in molecular OFF23 suggests that MACE-Osaka26 has robust generalization capabilities not only for periodic bulk crystals but also for isolated molecular systems.

\begin{table*}[t]
\centering
\caption{Performance comparison between MACE-Osaka24 and MACE-Osaka26. Accuracies for energy, force, and stress are evaluated. For MPtrj and OFF23, metrics are reported on both the training and test sets; for the HE26 dataset, metrics are on the training and validation sets. Benchmark accuracies are evaluated using two metrics: mean absolute error (MAE) and root mean square error (RMSE). The values outside and inside the parentheses represent the MAE and RMSE, respectively.}
\label{tab:1}
\begin{tabular}{lcccccc}
\hline 
Dataset & Split & \begin{tabular}[c]{@{}c@{}}\# \\ Structures\end{tabular} & Model & \begin{tabular}[c]{@{}c@{}}Energy MAE (RMSE)\\ (meV/atom)\end{tabular} & \begin{tabular}[c]{@{}c@{}}Force MAE (RMSE)\\ (meV/Å)\end{tabular} & \begin{tabular}[c]{@{}c@{}}Stress MAE (RMSE)\\ (meV/\AA$^3$)\end{tabular} \\ \hline
MPtrj & Train & 1,502,422 & MACE-Osaka24 & 31.4 (74.6) & 47.4 (\textbf{117.0}) & \textbf{1.7} (\textbf{10.3}) \\
(Crystalline) & & & MACE-Osaka26 & \textbf{26.8} (\textbf{60.5}) & \textbf{47.3} (148.3) & 1.7 (10.4) \\ 
 & Test & 77,973 & MACE-Osaka24 & 33.7 (77.2) & \textbf{61.9} (144.3) & 2.2 (45.8) \\
 & & & MACE-Osaka26 & \textbf{29.3} (\textbf{62.0})  & 62.4 \textbf{(140.7)} & \textbf{2.2} (\textbf{35.8}) \\ \hline
OFF23 & Train & 951,005 & MACE-Osaka24 & 5.7 (26.6) & \textbf{39.9} (\textbf{78.1}) & N/A \\
(Molecular) & & & MACE-Osaka26 & \textbf{4.9} (\textbf{11.9}) & 44.5 (88.0) & N/A \\ 
 & Test & 50,195 & MACE-Osaka24 & 5.8 (25.2) & \textbf{41.2} (\textbf{130.7}) & N/A \\
 & & & MACE-Osaka26 & \textbf{5.0} (\textbf{12.3}) & 45.7 (137.1) & N/A \\ \hline
HE26 & Train & 59,438 & MACE-Osaka24 & N/A\textsuperscript{a} & N/A\textsuperscript{a} & N/A\textsuperscript{a} \\
(Actinides etc.) & & & MACE-Osaka26 & \textbf{44.7} (\textbf{117.3}) & \textbf{26.3} (\textbf{73.7}) & \textbf{1.3} (\textbf{7.4}) \\ \hline 
\multicolumn{7}{l}{\textsuperscript{a} MACE-Osaka24 does not support the heavy elements included in HE26 dataset.}
\end{tabular}
\end{table*}

\subsection{\label{sec:results:A}Cross-domain Accuracy}

\begin{figure*}
\includegraphics[width=2\columnwidth]{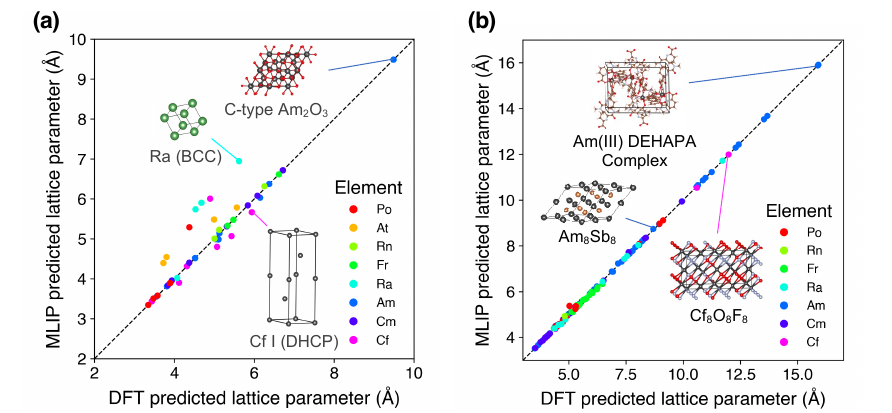}
\caption{\label{fig:3}Parity plots of DFT-calculated versus MLIP-predicted lattice parameters for heavy element systems. (a) The Basic Heavy Element (BHE) subset, comprising 42 unique structures. (b) The Complex Heavy Element (CHE) subset, comprising 133 unique structures. Three structures were excluded because their structural optimizations did not converge. The specific systems illustrated in the insets represent typical experimentally synthesized examples. 
The lattice parameter of each structure is defined throughout this paper as the average length of the unit cell lattice vectors.}
\end{figure*}

Furthermore, the results on the HE26 dataset underscore the model's expanded chemical coverage. While MACE-Osaka24 did not support the eight heavy elements (e.g., Am, Cm, and Cf) included in this dataset, MACE-Osaka26 successfully incorporates them, achieving an energy MAE of 44.7 meV/atom and an RMSE of 117.3 meV/atom on the training set (Figure~\ref{fig:2}(c)). Notably, the force and stress accuracies for this dataset are remarkably high (force MAE: 26.3 meV/\AA; stress MAE: 1.3 meV/\AA$^3$), confirming that the model effectively learns the complex potential energy surfaces of heavy-element compounds despite the limited dataset size.  These results demonstrate that MACE-Osaka26 is a universal potential, capable of providing reliable predictions across a diverse chemical space of 97-elements.

\subsection{\label{sec:results:B}Detailed Evaluation on HE26 Dataset}

Having established the broad cross-domain accuracy of MACE-Osaka26 spanning 97 elements, we now conduct a detailed evaluation focusing specifically on the HE26 dataset. Developing robust MLIPs for heavy elements---particularly actinides and lanthanides---presents a formidable challenge due to their intricate electronic structures and the inherent scarcity of reliable DFT training data. To systematically assess the model's predictive capabilities, physical fidelity, and generalization robustness, we analyze its performance across three subsets of increasing compositional and structural complexity: (1) elemental solids and binary oxides (BHE), (2) diverse multi-component systems (CHE), and (3) binary-to-quinary compositionally complex fluorite oxides (CCFO).

\subsubsection{\label{sec:results:B.1}Elemental Solids and Binary Oxides}

We evaluated the prediction accuracy of lattice parameters (defined as the average length of the unit cell lattice vectors) for the BHE subset of HE26 dataset, which consists of 42 unique structures. Figure~\ref{fig:3}(a) presents the parity plot comparing the lattice parameters predicted by MACE-Osaka26 against the DFT reference values. The model achieves an overall RMSE of 0.4488~\AA. The model demonstrates high accuracy for actinide elements such as Am and Cm, yielding near-zero errors for both elemental solids and oxides. Conversely, noticeable deviations are observed for specific structures containing Ra, At, and Cf. These deviations are largely attributed to the extreme difficulty in achieving electronic convergence in DFT calculations for these specific heavy elements, which inherently limits the amount of available training data. Nevertheless, the parity plot indicates that MACE-Osaka26 correctly captures the overall structural scale across diverse systems, confirming its viability as a solid baseline model even in regions with sparse reference data.

\subsubsection{\label{sec:results:B.2}Multi-Component Systems}

Building upon the baseline performance, we further assessed the model on the CHE subset of HE26 dataset. This dataset encompasses 136 unique structures and covers a remarkably broad chemical space comprising 64 different element types (excluding At). 
As shown in Figure~\ref{fig:3}(b), the MLIP predictions for the CHE subset are in good agreement with the DFT reference, achieving a very low RMSE of 0.0477~\AA.
This is an order of magnitude smaller than the 0.4488~\AA RMSE for the BHE subset shown in Figure~\ref{fig:3}(a).
Three structures were excluded from the analysis because their geometry optimization failed to converge. Remarkably, the model maintains this high accuracy across diverse chemical environments, including experimentally synthesized representative structures such as the Am(III) N,N-di-2-ethylhexyl-6-amide-pyridine-2-carboxylic acid (DEHAPA) complex~\cite{xu2023complexation} and Cf oxyfluoride \ce{Cf$_8$O$_8$F$_8$}\cite{peterson1968preparation} (illustrated as insets in Figure~\ref{fig:3}(b)). The comprehensive elemental coverage and composition of this subset are given in the Appendix (Tables~\ref{tab:4} and \ref{tab:5}). The consistently minimal errors across these 133 diverse structures demonstrate that MACE-Osaka26 successfully generalizes to multi-component heavy-element systems without experiencing performance degradation.

\subsubsection{\label{sec:results:B.3}Compositionally Complex Fluorite Oxides}

\begin{table}[b]
  \centering
  \caption{Error statistics of the MACE-Osaka26 model for predicting lattice parameters of compositionally complex fluorite oxides, categorized by the number of constituent elements (excluding oxygen). MaxAE denotes the maximum absolute error.}
  \label{tab:fluorite_error_statistics}
  \begin{tabular}{lcccc}
    \hline 
    \# Elements & \# Structures & RMSE (\AA) & MAE (\AA) & MaxAE (\AA) \\
    \hline
    1    & 17   & 0.0103 & 0.0071 & 0.0335 \\
    2    & 324  & 0.0142 & 0.0050 & 0.2090 \\
    3    & 1691 & 0.0158 & 0.0049 & 0.2223 \\
    4    & 1946 & 0.0170 & 0.0049 & 0.2333 \\
    \hline 
  \end{tabular}
\end{table}

Finally, we evaluated the model's performance on the CCFO subset, which encompasses solid solutions ranging from binary to quinary systems. A significant challenge for MLIPs is not merely minimizing the mathematical error against DFT, but accurately capturing the underlying physical and chemical trends across the periodic table.

\begin{figure*}[t]
\includegraphics[width=2\columnwidth]{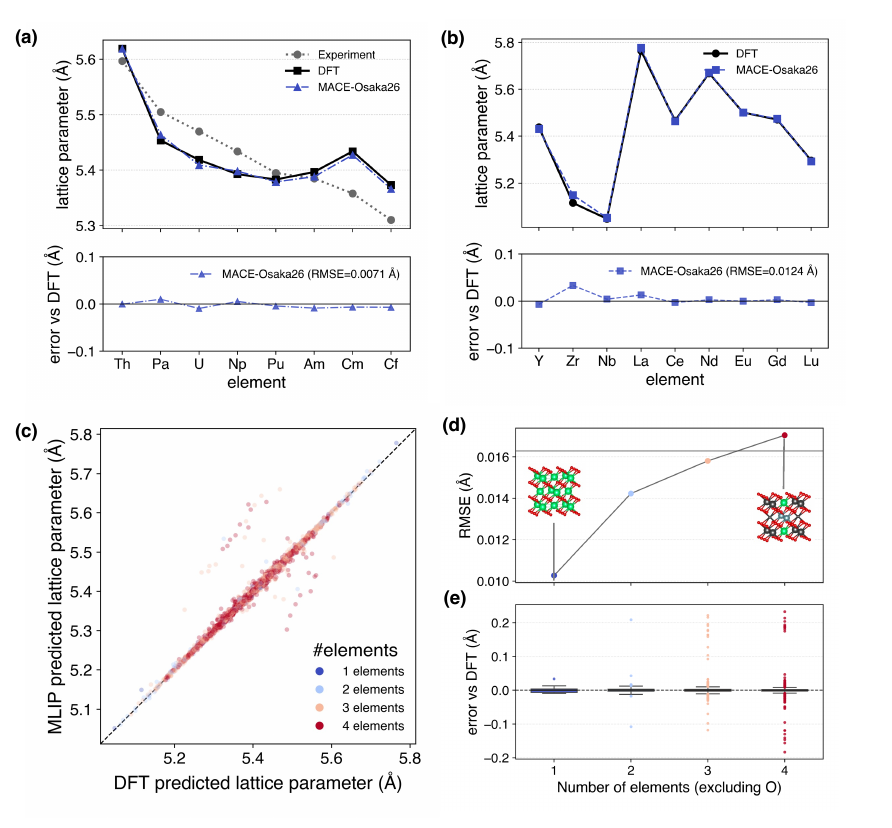}
\caption{\label{fig:4} Evaluation of MACE-Osaka26 on compositionally complex fluorite oxides. (a) Comparison of predicted lattice parameters with DFT reference values and experimental data across the actinide series (Th to Cf). The bottom panel shows the error of MACE-Osaka26 relative to DFT. (b) Predicted lattice parameters for fluorite-structured oxides containing various rare metals and lanthanides. (c) Parity plot of MLIP-predicted versus DFT-calculated lattice parameters, color-coded by the number of constituent elements (excluding oxygen). (d) Root-mean-square error (RMSE) and (e) error distribution relative to DFT as a function of the number of constituent elements.}
\end{figure*}

As shown in Figure~\ref{fig:4}(a), MACE-Osaka26 reproduces the DFT reference values across the actinide series (Th to Cf) with an RMSE of 0.0067~\AA. However, a noticeable discrepancy between the computational results and the experimental data appears around Am and Cm. While experiments show a monotonic decrease in the lattice constant associated with actinide contraction, the DFT data predict a slight increase at Cm. This deviation suggests that, although the MLIP properly learns the provided potential-energy surface, the PBE-based DFT approach adopted in this study may not fully describe the complex behavior of strongly correlated $f$ electrons in the heavier actinides. Nevertheless, the model's ability to closely follow the DFT trend supports its reliability as an emulator. Similarly, for fluorite-structured oxides containing various transition metals and lanthanides (Figure~\ref{fig:4}(b)), the model captures the structural variations with a low RMSE of 0.0124~\AA.

To further assess the model's robustness against compositional complexity, we analyzed the prediction accuracy as a function of the number of constituent elements (excluding oxygen). The parity plot in Figure~\ref{fig:4}(c) illustrates a tight correlation between the MLIP predictions and DFT calculations, irrespective of the system's complexity. The detailed error distributions are shown in Figure~\ref{fig:4}(d, e), and the quantitative metrics are summarized in Table~\ref{tab:fluorite_error_statistics}. Although the RMSE slightly increases from 0.0103~\AA\ for binary systems to 0.0170~\AA\ for quinary systems, the mean absolute error (MAE) remains exceptionally low ($\sim$0.005~\AA) and highly consistent across all complexity levels. These findings confirm that MACE-Osaka26 maintains high predictive accuracy and strong generalization capabilities even for highly complex, multi-component heavy-element ceramics.

\begin{figure*}[t]
\includegraphics[width=2\columnwidth]{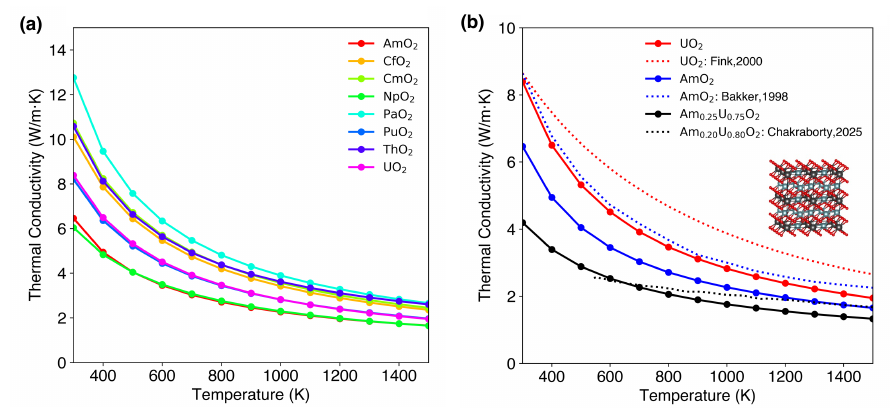}
\caption{\label{fig:5} Lattice thermal conductivity ($\kappa_L$) of actinide oxides predicted by MACE-Osaka26. (a) Temperature dependence of $\kappa_L$ for a series of actinide dioxides (\ce{$An$O_2}, $An$ = Th, Pa, U, Np, Pu, Am, Cm, Cf). The solid lines with symbols represent the MLIP results calculated by solving the Wigner transport equation via phono3py, while the open symbols denote experimental data from the literature where available. (b) Comparison of $\kappa_L$ between the pure binary oxides and the ternary solid solution \ce{Am_{0.25}U_{0.75}O_2}. The experimentally obtained lattice thermal conductivities for pure \ce{UO_2}~\cite{fink2000uo2}, \ce{AmO_2}~\cite{bakker1998thermal}, and the solid solution \ce{Am_{0.2}U_{0.8}O_2}~\cite{chakraborty2025thermal} are plotted as dotted curves to validate the MLIP predictions.}
\end{figure*}

\subsection{\label{sec:results:C}Lattice Thermal Conductivity}

Finally, to demonstrate the applicability of MACE-Osaka26 to demanding thermal properties, we evaluated the lattice thermal conductivity ($\kappa_L$) of actinide fluorite oxides. The calculations were performed by solving the phonon Wigner transport equation~\cite{NatPhys.15.809,PhysRevX.12.041011}, leveraging the second- and third-order interatomic force constants generated by the MACE-Osaka26 MLIP.

Figure~\ref{fig:5}(a) presents the predicted temperature dependence of $\kappa_L$ for a comprehensive series of actinide dioxides (\ce{$An$O_2}, where $An$ = Th, Pa, U, Np, Pu, Am, Cm, and Cf). For systems where reference experimental data are available (such as \ce{UO_2}~\cite{fink2000uo2}, \ce{AmO_2}~\cite{bakker1998thermal}, \ce{ThO_2}~\cite{malakkal2019thermal}, \ce{PuO_2}~\cite{wang2015thermal,cozzo2011thermal,gibby1971effect}, \ce{CmO_2}~\cite{konings2001thermochemical} and \ce{NpO_2}~\cite{nishi2008thermal}), the absolute values are consistent with experimental reports across a wide temperature range (300--1500~K).  It is particularly noteworthy that a single universal model successfully evaluated the thermal properties across this entire actinide series without yielding any imaginary phonon frequencies. The complete absence of imaginary modes across all eight distinct systems confirms the robust dynamical stability of the predicted structures and underscores the high physical fidelity of the learned anharmonic potential energy surface for heavy elements.

To investigate the impact of transuranium element doping---a critical scenario for advanced nuclear fuel engineering and waste transmutation---we extended this analysis to a mixed oxide system. The solid solution was systematically modeled by substituting one U atom with an Am atom within the 12-atom conventional fluorite unit cell (\ce{U_4O_8} $\rightarrow$ \ce{U_3Am_1O_8}), yielding \ce{Am_{0.25}U_{0.75}O_2}. Crucially, as shown in Figure~\ref{fig:5}(b), we compared these MLIP predictions against experimentally obtained lattice thermal conductivities for pure \ce{UO_2}~\cite{fink2000uo2}, pure \ce{AmO_2}~\cite{bakker1998thermal}, and the \ce{Am_{0.20}U_{0.80}O_2}~\cite{chakraborty2025thermal} solid solution. Despite the slight difference in Am concentration dictated by the periodic boundary conditions of the unit cell (25\% in the model versus 20\% in the experiment), the MLIP quantitatively captures the severe degradation of thermal performance induced by Am doping, seamlessly bridging the properties of the two pure end-members. 

\section{\label{sec:discussion}Discussion}

The superior performance of MACE-Osaka26, characterized by reduced outliers and improved stability in heavy element systems, is largely attributed to the increased graph cutoff radius. MACE-Osaka24 employed a shorter cutoff of $4.5\text{ \AA}$, which was previously reported to be insufficient for accurately describing heavy element BCC structures due to the limited neighbor lists (as discussed in the Appendix of Ref.~\cite{Shiota2025MACEOsaka24TEA}). By extending the cutoff to $6.0\text{ \AA}$ in MACE-Osaka26, the model can capture longer-range interactions essential for these systems. This modification not only resolves the issues with heavy elements but also enhances the generalization capability across diverse chemical environments, including molecular systems (OFF23) and general crystalline structures (MPtrj).

The prediction accuracy for the lattice parameters of the BHE subset, which contains elemental solids and binary oxides, was found to be an order of magnitude lower than that of the CHE subset, which includes complex multi-component systems. This discrepancy is primarily attributed to the smaller sample size of the BHE subset (42 structures) compared to the CHE subset (136 structures), as well as the presence of specific outliers with errors approaching 1.0 Å. Except for a single case of Cf, the elements exhibiting large deviations are not included in the relatively larger CCFO subset, indicating a lower representation in the training data. These results suggest that further expansion of the dataset will likely lead to improvements in prediction accuracy for these heavy-element systems.

In the lattice thermal conductivity predictions for actinide FOs, the total absence of imaginary phonon modes across the entire actinide series is particularly significant, as it demonstrates the robust dynamical stability of the predicted structures and the high physical fidelity of the learned anharmonic potential energy surface. Furthermore, extracting third-order force constants for disordered systems typically requires extremely large supercells, making conventional DFT calculations computationally prohibitive. The ability of MACE-Osaka26 to efficiently and accurately model these thermophysical properties of actinide FOs indicates its suitability for the computational design of next-generation nuclear fuels.

\section{\label{sec:conclusion}Conclusion}

In this study, we constructed a heavy element first-principles calculation dataset HE26. Existing large-scale datasets such as MPtrj cover up to 89 elements, but for certain heavy elements outside this range, essentially no first-principles dataset has been available. HE26 fills this gap by explicitly including eight heavy elements (i.e., Am, Cm, Cf, Fr, At, Ra, Po, and Rn). The dataset spans elemental solids, binary oxides, multi-component compounds drawn from the literature and crystallographic databases, and compositionally complex actinide--lanthanide fluorite-type solid solutions. By integrating HE26 with the molecular (OFF23) and crystalline (MPtrj) datasets through the total energy alignment (TEA) protocol, we developed MACE-Osaka26, a universal MLIP covering 97 elements. To the best of our knowledge, this is the broadest elemental coverage among existing MLIPs. The model and data are both publicly freely available, providing a reproducible starting point for building universal MLIPs that include heavy elements.

The resulting MACE-Osaka26 maintains accuracy and generalization across the molecular and crystalline domains. The model also stably reproduces property trends along structural scales and compositional series for the heavy element systems covered by HE26. Furthermore, we evaluated the lattice thermal conductivity of actinide fluorite oxides and their solid solutions. Second- and third-order interatomic force constants were generated with the MLIP, and the lattice thermal conductivity $\kappa_L$ was obtained by solving the Wigner transport equation. The results agreed with reference experiments over a wide temperature range. For a transuranium-doped solid solution such as $\text{Am}_{0.25}\text{U}_{0.75}\text{O}_2$, the model quantitatively captured the large reduction in thermal conductivity caused by Am doping. These results demonstrate that a universal MLIP can provide a scalable route to evaluating thermal transport properties of compositionally complex systems containing transuranium elements. Such evaluations have been not easy to carry out with first-principles calculations alone.

However, HE26 and MACE-Osaka26 represent only a first step towards accelerating computation in the heavy element regime. The reference electronic structures in this study are based on PBE (and partially DFT+U), chosen for compatibility with MPtrj. For transuranium element systems where strongly correlated $f$-electrons, magnetism, and spin--orbit coupling (SOC) are significant, the validity of the reference calculations itself can become a source of uncertainty. Going forward, it will be important to (i) generate datasets based on electronic structure calculations that include SOC, (ii) cross-check with higher-level electronic structure methods, (iii) expand external validation against experimental data (lattice constants, thermochemistry, elasticity, thermal transport), and (iv) strategically increase data through uncertainty quantification and active learning. Through these efforts, the reliability and scale of datasets covering transuranium elements should be improved step by step.

Today's MLIPs have reached practical utility by starting from relatively small initial datasets and building up through continuous data expansion and validation. HE26 serves as a starting point for extending the same development path into the heavy element regime. Moreover, the intermediate features learned by universal MLIPs can serve as compact atomic descriptors for chemical properties beyond energy and its derivatives~\cite{shiota2024universal,doi:10.1139/cjc-2023-0152, gouvea2026combining, kim2025leveraging, bojan2026representing}. Extending these descriptors to 97 elements broadens the accessible chemical space for materials design and screening through materials informatics.
In summary, this work my provide an open foundation for accelerating large-scale screening and property prediction of heavy elements in nuclear materials science, encompassing nuclear fuels, waste forms, and actinide-bearing high-entropy ceramics.

\section{\label{sec:methods}Methods}

\subsection{\label{sec:method:A}HE26 Dataset}
The HE26 dataset was generated using the Vienna \textit{Ab initio} Simulation Package (VASP) \cite{kresse1993ab,kresse1996efficient,kresse1994ab,kresse1996efficiency} and extends the elemental coverage to eight heavy elements: Am, Cm, Cf, Fr, At, Ra, Po, and Rn. 
For actinide-containing systems, where electronic-structure convergence can be sensitive to magnetism, we first identified a suitable magnetic configuration via single-point calculations over multiple candidate magnetic states. Details of this magnetic-structure selection procedure are provided in Appendix~\ref{sec:appendix:A.1.b}. Geometry optimizations were then performed starting from the selected configuration.

After the DFT calculations, we curated the raw HE26 structures to improve data quality and reduce noise during training. Specifically, we removed (i) structures whose geometry optimizations did not complete, (ii) structures with positive total energies for all elements except the noble gas Rn (to avoid spurious high-energy outliers), and (iii) structures for which the self-consistent field (SCF) iterations failed to converge during geometry optimization.

In the following subsections, we describe the generation procedures for each HE26 subset: BHE, CHE, and CCFO.

\subsubsection{\label{sec:method:A.1}Basic Heavy Element Subset}

For the elemental systems and binary oxides of eight heavy elements (Am, Cm, Cf, Fr, At, Ra, Po, and Rn), we compiled structural data using a stepwise approach. First, we utilized available Crystallographic Information File (CIF) formats from the Crystallography Open Database (COD) database~\cite{Grazulis2009COD,Grazulis2012CODNAR}. 
For structures not included in the COD, initial geometries were generated by searching the literature. 
Additionally, for basic crystal structures—specifically body-centered cubic (bcc), face-centered cubic (fcc), and hexagonal close-packed (hcp)—that could not be obtained from experimental values, initial geometries were generated using a same-volume approximation derived from the experimental values of other structures. 

\subsubsection{\label{sec:method:A.2}Complex Heavy Element Subset}

For the complex multi-element systems of eight heavy elements not found in the MPtrj dataset, we compiled structural data from multiple sources. 
First, we used available CIF files from the COD database. 
Any equivalent structures were excluded using \texttt{pymatgen}'s \texttt{StructureMatcher }~\cite{Ong2013pymatgen}, even if they possessed different COD ID.  Additionally, for structures not registered in the COD database, we searched the literature as extensively as possible and prepared the corresponding initial structures.
To further increase the number of structures, we retrieved additional structures from the Inorganic Substances Database book 2017 \cite{VillarsCenzualGladyshevskii+2017+1671+1956} that were not covered by the previous steps and generated their corresponding initial geometries.

\subsubsection{\label{sec:method:A.3}
Compositionally Complex Fluorite Oxide Subset} The initial lattice constant of fluorite \ce{Th4O8} (the conventional unit cell of fluorite \ce{ThO2}) was obtained from MPtrj. Subsequently, initial structures prior to geometry optimization were generated for all combinations of 1--4 substitutions of Th with $X \in$ \{{La, Ce, Nd, Eu, Gd, Lu, Y, Zr, Nb, Th, Pa, U, Np, Pu, Am, Cm, Cf}\}. \ce{Th4O8} was selected as the parent cell because it possesses the largest lattice constant within the \ce{$An$O2} (An = Th, Pa, U, Np, Pu, Am, Cm, Cf) series \cite{prodan2007_covalency_actinide_dioxides_hse, benedict1987_structural_data_actinide_elements_binary_compounds, wang2020_water_actinide_dioxide_surfaces_review}. Starting with the largest host lattice minimizes the risk of unphysically short interatomic distances in the initial substituted structures prior to relaxation. This approach ensures that the dataset robustly covers compositionally complex mixed oxides containing not only actinides but also critical minor metal elements and transition metals.The actinide subset (Th, Pa, U, Np, Pu, Am, Cm, Cf) was chosen because stable bulk actinide dioxides crystallizing in the \ce{CaF2}-type (fluorite) structure have been experimentally established across this series \cite{prodan2007_covalency_actinide_dioxides_hse}. Actinium was excluded because its chemistry is strongly dominated by the +III oxidation state, and a stable fluorite \ce{AcO2} phase has not been observed \cite{deblonde2021_coordination_properties_ionic_radius_actinium, fried1950_preparation_identification_pure_actinium_compounds}. To explicitly encompass rare-earth elements, yttrium (Y) and a representative lanthanide subset (La, Ce, Nd, Eu, Gd, Lu) were included. These rare-earth species are technologically critical as abundant nuclear fission products and common aliovalent dopants. They frequently form substituted fluorite solid solutions, such as \ce{Y2O3}-doped \ce{ThO2} and various \ce{UO2}-based matrices. This selection systematically spans a broad range of ionic radii across the lanthanide contraction while simultaneously sampling diverse 4$f$-electron configurations and redox flexibility \cite{vinograd2021_thermodynamic_structural_modelling_lndoped_uo2, kim2008_applicability_ceo2_surrogate_puo2_mox, alexandrov2010_defect_clustering_tho2_trivalent_oxides}. Finally, transition-metal cations (Zr, Nb) were selected because fluorite-type actinide dioxide matrices commonly accommodate them as dopants or within macroscopic solid solutions. For instance, \ce{UO2}--\ce{ZrO2} and \ce{ThO2}--\ce{ZrO2} readily form extended fluorite solid solutions, and fluorite \ce{UO2} exhibits measurable Nb solubility upon \ce{Nb2O5} addition \cite{STAICU20156,zhang2017_thermochemistry_uo2_tho2_uo2_zro2_fluorite_solid_solutions, harada1996_sintering_niobia_doped_uo2_pellet}.

\subsection{\label{sec:method:B}97-element Universal Dataset}

In Ref.~\cite{Shiota2025MACEOsaka24TEA}, we integrated the MPtrj and OFF23 datasets using TEA protocol. It should be noted that the dispersion correction values were subtracted from the OFF23 dataset. Following the data splitting protocol established in Ref.~\cite{batatia2024foundationmodelatomisticmaterials} and Ref.~\cite{kovacs2023mace}, 95\% of the MPtrj and OFF23 datasets were used for training and validation, while the remaining 5\% were reserved for testing. The HE26 dataset was generated under conditions compatible with the MPtrj dataset, allowing for its direct integration into the MPtrj+OFF23 dataset. Although the HE26 dataset encompasses nearly all known experimentally synthesized crystal and complex structures, its volume is small compared to MPtrj and OFF23. Therefore, all data in HE26 were used for training and validation without a separate test split. In this study, the combined training and validation data are referred to as the ``training set." We designate this combined MPtrj+OFF23+HE26 dataset as the ``97-element universal dataset."

\subsection{\label{sec:method:C}MLIP Training}

Using the 97-element universal dataset (MPtrj + OFF23 + HE26), we trained MLIPs with the MACE framework~\cite{batatia2022mace,batatia2024foundationmodelatomisticmaterials}, using mace v0.3.12 (\url{https://github.com/ACEsuit/mace}). 
We denote the resulting MLIP as MACE-Osaka26. 
The model and the final training data are available at \url{https://github.com/qiqb-osaka/mace-osaka26}.
Training followed the hyperparameters, cost functions, and optimizers of the MACE-MP-0 described in Ref.~\cite{batatia2024foundationmodelatomisticmaterials}, with minor modifications. 
For this model, we set a cutoff radius of 6.0 ~\AA{} for constructing the atomic neighborhood graph.
We adopted the same atomic reference energies as MACE-MPA-0 for the 89 elements present in MPtrj, while calculating spin-polarized VASP energies for the isolated atoms of the eight newly added heavy elements.
We incorporate a short-range repulsive correction via the Ziegler–Biersack–Littmark (ZBL)~\cite{ziegler1985stopping} screened nuclear potential and, to make it consistent across elements, apply an element-dependent Agnesi distance transformation~\cite{witt2023acepotentials} based on covalent radii before evaluating the radial basis.
The model was trained using 32 NVIDIA A100 GPUs across 4 nodes on SQUID, the large-scale computing system at D3 Center, The University of Osaka.

\subsection{\label{sec:method:C}Lattice Thermal Conductivity Calculation}

Although lattice thermal conductivity can in principle be evaluated using the Green--Kubo formalism, lattice thermal conductivity in crystalline solids arising from anharmonic phonon--phonon interactions can be computed more efficiently within a lattice-dynamics framework by solving phonon transport equations, such as the Peierls--Boltzmann transport equation (PBTE) or the Wigner transport equation (WTE) \cite{Green1954GK,Kubo1957,NatPhys.15.809,PhysRevX.12.041011}. 
The predictive accuracy of these approaches depends critically on the fidelity of the underlying interatomic forces. 
While empirical force fields are computationally efficient, their transferability is often limited, whereas first-principles calculations provide more accurate interatomic forces at a substantially higher computational cost. 
In particular, the calculation of third-order anharmonic interatomic force constants (IFCs) using DFT becomes a major bottleneck for multicomponent, low-symmetry, or defect-containing systems, because the required supercell sizes and the number of displaced configurations increase rapidly as the crystal symmetry is reduced \cite{Togo2023JPSJ}. 
To balance accuracy and computational efficiency, the lattice thermal conductivity ($\kappa_L$) and phonon dispersion relations in this study were calculated using the finite displacement method implemented in the \texttt{phono3py} package \cite{phono3py,phonopy-phono3py-JPCM}, with atomic forces evaluated using the trained MACE-Osaka26 potential.

For certain heavy elements (e.g., Am) whose default atomic masses are not registered in \texttt{phono3py}, standard atomic weights were specified manually. Prior to the phonon calculations, full geometry optimizations of all structural models were performed until the residual forces on each atom converged to less than $0.03$~eV/\AA.

For the pure actinide dioxides (\ce{$An$O_2}), we adopted a supercell containing a total of 96 atoms, constructed by expanding the 12-atom fluorite conventional unit cell to $2 \times 2 \times 2$. The second- and third-order interatomic force constants (IFCs) were extracted using a finite atomic displacement distance of $0.03$~\AA. To model the mixed oxide solid solution (\ce{Am_{0.25}U_{0.75}O_2}), one U atom within the 12-atom conventional unit cell was substituted with an Am atom (\ce{U_4O_8} $\rightarrow$ \ce{U_3Am_1O_8}), which was subsequently expanded into a $2 \times 2 \times 2$ supercell. In this mixed system, the reduction in symmetry leads to a massive proliferation in the number of force evaluations (displacement patterns) required to extract the third-order IFCs. Therefore, to keep the computational cost within a practical range, a cutoff radius of 6~\AA\ was applied to the IFC calculations. 

Crucially, the macroscopic heat transport was evaluated by solving the phonon Wigner transport equation rather than the conventional Peierls-Boltzmann transport equation (PBTE). Under this Wigner formalism---formulated by Simoncelli et al.\ and Isaeva et al.\ and implemented in \texttt{phono3py}---the total lattice thermal conductivity tensor $\kappa_L^{\alpha\beta}$ is described as the sum of two components:
\begin{equation}
    \kappa_L^{\alpha\beta} = \kappa_P^{\alpha\beta} + \kappa_C^{\alpha\beta}
\end{equation}
Here, $\alpha$ and $\beta$ are subscripts denoting the Cartesian coordinate directions ($x, y, z$). The first term, $\kappa_P^{\alpha\beta}$, originates from the diagonal elements of the heat flux operator and represents the conventional PBTE contribution (particle-like heat transport) governed by phonon group velocities and relaxation times. Conversely, the second term, $\kappa_C^{\alpha\beta}$, arises from the off-diagonal elements and captures the ``wave-like coherent heat transport'' (inter-band tunneling), reflecting the coupling between different phonon branches ($j \neq j^\prime$) with proximate angular frequencies. Incorporating this coherent component, $\kappa_C^{\alpha\beta}$, is theoretically indispensable for the low-symmetry \ce{Am_{0.25}U_{0.75}O_2} supercell. Due to zone folding in this structure, optical phonon modes densely proliferate and intertwine, reducing the inter-band energy differences to a magnitude comparable to the phonon linewidths (scattering rates).

The macroscopic scalar lattice thermal conductivity, $\kappa_L$, reported in this study is defined as the average of the three diagonal components of the tensor:
\begin{equation}
    \kappa_L = \frac{1}{3} (\kappa_L^{xx} + \kappa_L^{yy} + \kappa_L^{zz})
\end{equation}
The Brillouin zone integration for this evaluation was performed using a uniform $10 \times 10 \times 10$ $q$-point mesh. In the present calculations, isotope scattering was excluded to isolate and focus on the intrinsic anharmonic phonon-phonon scattering, as well as the point defect scattering originating from mass and force-constant variances in the mixed oxide system.

\begin{acknowledgments}
This project was supported by funding from the MEXT Quantum Leap Flagship Program (MEXTQLEAP) through Grant No. JPMXS0120319794, the JST COI-NEXT Program through Grant No. JPMJPF2014, and the JST ASPIRE Program Grant No. JPMJAP2319. The completion of this research was partially facilitated by the JSPS Grants-in-Aid for Scientific Research (KAKENHI), specifically Grant Nos. JP23H03819, and JP25K23358. We thank the Supercomputer Center, the Institute for Solid State Physics, the University of Tokyo, for allowing us to use their facilities. A part of the calculations were performed using the Genkai supercomputer of the Research Institute for Information Technology at Kyushu University. This work was also achieved using the SQUID supercomputer at the Cybermedia Center, The University of Osaka.
\end{acknowledgments}

\appendix

\begin{figure*}[t]
\includegraphics[width=2\columnwidth]{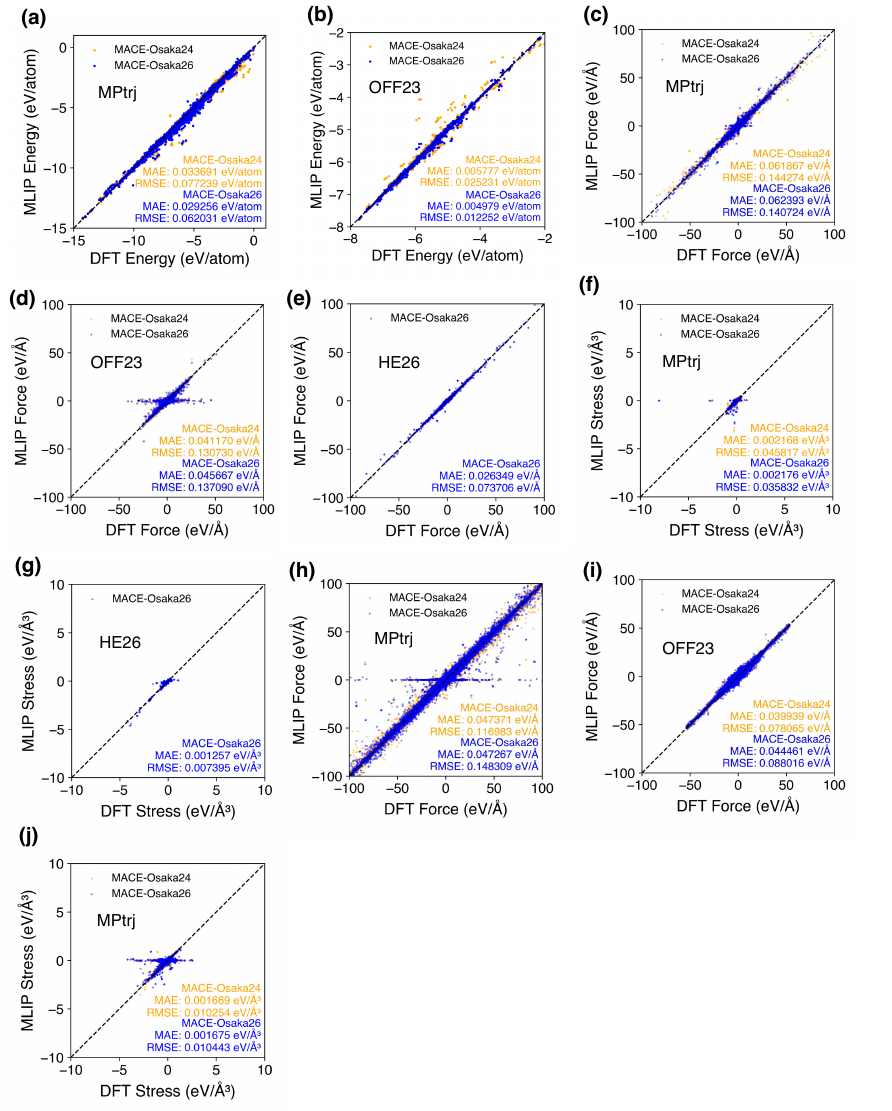}
\caption{\label{fig:6} Accuracy comparison of energies, atomic forces, and virial stresses between MACE-Osaka24 and MACE-Osaka26. (a, b) Parity plots for energies evaluated on (a) the MPtrj test set and (b) the OFF23 test set. (c–e) Parity plots for atomic forces evaluated on (c) the MPtrj test set, (d) the OFF23 test set, and (e) the Heavy (HE26) dataset. (f, g) Parity plots for virial stresses evaluated on (f) the MPtrj test set and (g) the Heavy (HE26) dataset. (h–j) Parity plots for training set evaluated for atomic forces on (h) the MPtrj test set and (i) the OFF23 test set, and for virial stresses on (j) the MPtrj test set. Note that for the Heavy (HE26) dataset where MACE-Osaka24 is not applicable or not supported, only the results for MACE-Osaka26 (blue) are shown. Mean absolute errors (MAE) and root-mean-square errors (RMSE) are provided in units of eV/atom for energies, eV/Å for forces, and eV/Å$^3$ for stresses.}
\end{figure*}

\begin{figure*}
\includegraphics[width=2\columnwidth]{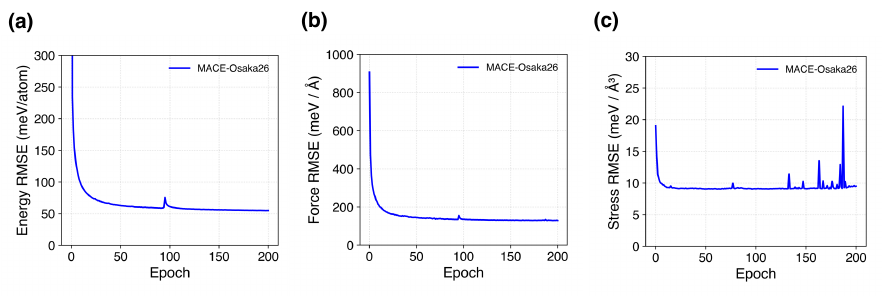}

\caption{\label{fig:7} Root mean square errors (RMSEs) obtained during training of the MACE‑Osaka26 model for 200 epochs. (a) RMSE of the energy per atom (meV/atom). (b) RMSE of the forces (meV/Å). (c) RMSE of the stresses (meV/Å³).}
\end{figure*}

\section{\label{sec:appendix:A}Computational Details}

\subsection{\label{sec:appendix:A.1}Electronic-structure reference calculations}

\subsubsection{\label{sec:appendix:A.1.a}General DFT Setting}

VASP calculations were performed using periodic boundary conditions and PAW pseudopotentials. Unless stated otherwise, the input settings followed the Materials Project-style defaults as implemented in \texttt{pymatgen}'s \texttt{MPRelaxSet}~\cite{Ong2013pymatgen}. We applied specific explicit overrides across all calculations to ensure consistency with recent large-scale dataset practices (e.g., OMat24~\cite{BarrosoLuque2024OMat24}). Specifically, we utilized the normal electronic minimization algorithm (\texttt{ALGO = Normal}) and set the electronic self-consistency convergence criterion to $10^{-5}$ eV (\texttt{EDIFF = 1e-05}). We applied Gaussian smearing (\texttt{ISMEAR = 0}) with a broadening width of 0.05 eV (\texttt{SIGMA = 0.05}), and the baseline charge density mixing parameters were set to \texttt{AMIX = 0.2} and \texttt{BMIX = 0.0001}. To improve computational efficiency and reduce I/O overhead, parallelization over bands was specified (\texttt{NPAR = 4}), and the writing of charge density and wavefunctions was suppressed (\texttt{LCHARG = .FALSE.}, \texttt{LWAVE = .FALSE.}). Other general parameters (e.g., $k$-point meshes) were strictly inherited from the \texttt{MPRelaxSet} defaults.

As a consequence of these inherited defaults and explicit settings, the calculations employed a plane-wave energy cutoff of 520.0 eV (\texttt{ENCUT = 520.0}), ``Accurate'' precision (\texttt{PREC = Accurate}), and automatic real-space projection (\texttt{LREAL = Auto}). To ensure the accurate treatment of localized states, non-spherical contributions from the PAW spheres were included (\texttt{LASPH = .TRUE.}), and the maximum angular momentum for charge density mixing was set to \texttt{LMAXMIX = 6}. Furthermore, though the writing of charge density and wavefunctions was suppressed, DOS and band structure projection data were still calculated and written (\texttt{LORBIT = 11}). Finally, the maximum number of electronic steps (\texttt{NELM}) was set to 100, extending to 1000 if convergence was not initially reached.

\subsubsection{\label{sec:appendix:A.1.b}Magnetic-configuration screening}

Various initial magnetic configurations were screened by single-point calculations at fixed ionic positions, explicitly disabling ionic relaxation by setting \texttt{IBRION = -1} and \texttt{NSW = 0}. The evaluated configurations included ferromagnetic (FM), antiferromagnetic (AFM), no spin, and the \texttt{pymatgen} default for the \texttt{MAGMOM} parameters (initial on-site magnetic moments of each atom). All other electronic minimization and I/O parameters followed the general settings described in Section~\ref{sec:appendix:A.1.a}. The magnetic configuration used for subsequent geometry optimizations was selected using an automated decision rule based on the resulting energies and Self-Consistent Field (SCF) convergence behavior. Specifically, we (i) considered only calculations for which the electronic self-consistency cycle converged normally; (ii) identified the minimum-energy-per-atom solution, treated solutions within 10 meV/atom of this minimum as energetically degenerate, and selected from this degenerate set the configuration with the smallest SCF iteration count; (iii) if the preliminary selection required more than 200 SCF iterations, we allowed a fallback to an alternative configuration within 20 meV/atom whose SCF iteration count was reduced by at least 100, to improve robustness; and (iv) additionally performed a non-spin-polarized single-point calculation with \texttt{ISPIN = 1}, and if its energy per atom was within 10 meV/atom of the minimum-energy solution, we selected this non-magnetic state and, for subsequent structural relaxations, kept \texttt{ISPIN = 2} while initializing all site moments to zero (i.e., \texttt{MAGMOM = 0} for all atoms). Finally, when a spin-polarized configuration was selected, the site-resolved converged magnetic moments of that run were used as the initial \texttt{MAGMOM} for subsequent structural relaxations.

\subsubsection{\label{sec:appendix:A.1.c}Geometry optimization}
Structural relaxations utilized the converged site-resolved magnetic moments from the initial screening step as the starting \texttt{MAGMOM} values. The geometry optimizations allowed for the full relaxation of atomic positions, cell shape, and volume (\texttt{ISIF = 3}) using the conjugate gradient algorithm (\texttt{IBRION = 2}). The maximum number of ionic steps (\texttt{NSW}) was initially set to 99; if convergence was not reached, this limit was extended to 1000 steps. The ionic convergence criterion was defined by a force threshold of -0.03 eV/\AA\ (\texttt{EDIFFG = -0.03}). All other electronic and general parameters were inherited from Section~\ref{sec:appendix:A.1.a}.

Exchange–correlation effects were treated within the generalized gradient approximation
using the Perdew–Burke–Ernzerhof (PBE) functional. For oxide and fluoride materials
containing Co, Cr, Fe, Mn, Mo, Ni, V, or W, Hubbard U corrections were applied
following the Materials Project defaults (pymatgen MPRelaxSet).

\section{\label{sec:appendix:B}Benchmarking Force and Stress on Cross-domain Datasets}

To complement the energy accuracy discussions in the main text, we provide a detailed evaluation of the total energies, atomic forces, and stresses predicted by MACE-Osaka26. Figure~\ref{fig:6} displays the parity plots for these properties across the MPtrj, OFF23, and HE26 dataset. The results confirm that the model maintains high predictive performance for derivative physical quantities, ensuring its applicability to diverse tasks such as structural optimization and molecular dynamics simulations.

\section{\label{sec:appendix:C}Details of HE26 Dataset}

Table \ref{tab:3} summarizes the information regarding the structural sources for the Basic Heavy Element (BHE) subset within the HE26 dataset. Tables \ref{tab:4} and \ref{tab:5} outline the methods used for the structural sources of the Complex Heavy Element (CHE) subset.

\section{\label{sec:appendix:D}Training Curves of MACE-Osaka26}

Figure \ref{fig:7} illustrates the evolution of validation accuracy for energy, forces, and stress during the 200-epoch training of the MACE-Osaka26 model. As the number of epochs increases, the errors in energy and forces decrease smoothly. Regarding the stress component, although a significant transient increase is observed during training, the values eventually converge to a nearly constant level. Since this study focuses solely on training the model corresponding to MACE-Osaka24-small, it is expected that adopting hyperparameters consistent with the "large" version—as seen in MACE-Osaka24—would yield a smoother convergence curve for stress.

\bibliography{apssamp}

\begin{verbatim}
\end{verbatim}

\afterpage{
\begin{widetext}
\begingroup\footnotesize
\setlength{\tabcolsep}{3pt}

\begin{table*}[t]
\centering
\caption{Literature and database sources for initial Basic Heavy Element (BHE) subset structures.}
\label{tab:3}
\footnotesize
\setlength{\tabcolsep}{3pt}
\begin{tabular}{llllHll}
\toprule
COD ID & Primitive Formula & Structure type & Crystal System & & Space Group No. & Ref. \\
\midrule
9008524 & \ce{Am} & dhcp & hexagonal & Am\_dhcp\_P63mmc\_exp-COD9008524-Wyckoff1963.cif & 194 & \cite{wyckoff1963crystal} \\
- & \ce{Am} &fddd & orthorhombic & Am\_AmIII\_fddd\_A10.115\_B5.670\_C3.116\_P10p9GPa\_exp-Heathman2000.cif & 70 & \cite{heathman_pressure_2000} \\
- & \ce{Am} & hcp & hexagonal & Am\_hcp\_P63mmc\_a3.4585\_c5.6477\_eqvol\_from\_dhcp\_a3.467\_c11.24\_ca1.6330.cif & 194 & This work \\
- & \ce{Am} & Pnma & orthorhombic & Am\_AmIV\_Pnma\_A5.093\_B4.679\_C3.028\_x0.403\_z0.101\_P17p6GPa\_exp-Heathman2000.cif & 62 & \cite{heathman_pressure_2000}\\
- & \ce{Am2O3} & A-type & trigonal & Am2O3\_Atype\_P-3m1.cif & 164 & \cite{horlait_self-irradiation_2014} \\
4124688 & \ce{Am2O3} & C-type & cubic & Am2O3\_Ctype\_Ia-3\_COD4124688.cif & 206 & \cite{templeton1953crystal} \\
9008998 & \ce{AmO2} & fluorite & cubic & AmO2\_Fm-3m\_COD9008998.cif & 225 &  \cite{wyckoff1963crystal}\\
- & \ce{At} & bcc & cubic & At\_bcc\_Im-3m\_a4.4181\_PRL2013\_SOC\_fromV0.cif & 229 & \cite{hermann_condensed_2013} \\
- & \ce{At} & bct & tetragonal & At\_bct\_I4mmm\_a3.8100\_c8.7630\_PRL2013\_SOC.cif & 139 & \cite{hermann_condensed_2013} \\
- & \ce{At} & fcc & cubic & At\_fcc\_Fm-3m\_a5.3882\_PRL2013\_SOC+D2.cif & 225 & \cite{hermann_condensed_2013}\\
- & \ce{At} & hcp & hexagonal & At\_hcp\_P63mmc\_a3.8179\_c6.2345\_PRL2013\_scalar\_fromV0.cif & 194 & \cite{hermann_condensed_2013} \\
- & \ce{Cf} & bcc & cubic & Cf\_bcc\_Im-3m\_a3.7923\_fromV0\_eqvol.cif & 229 & This work \\
- & \ce{Cf} & dhcp & hexagonal & Cf\_I\_dhcp\_P63mmc\_a3.3800\_c11.0250\_PRB2013.cif & 194 & \cite{heathman_structural_2013} \\
- & \ce{Cf} & fcc & cubic & Cf\_II\_fcc\_Fm-3m\_a4.7830\_trivalent\_PRB2013.cif & 225 & \cite{heathman_structural_2013} \\
- & \ce{Cf} & fddd & orthorhombic & Cf\_III\_fddd\_a8.4530\_b4.8910\_c2.6050\_\textasciitilde{}100GPa\_PRB2013.cif & 70 & \cite{heathman_structural_2013} \\
- & \ce{Cf} & hcp & hexagonal & Cf\_hcp\_P63mmc\_a3.3786\_c5.5172\_eqvol\_hypothetical.cif & 194 & This work \\
- & \ce{CfO} & NaCl & cubic & CfO\_Fm-3m\_a4.970\_theory-PRB2010\_SIC-LSD.cif & 225 & \cite{PhysRevB.81.045108} \\
- & \ce{CfO2} &  fluorite & cubic & CfO2\_Fm-3m\_a5.310\_exp.cif & 225 & \cite{baybarz_californium_1972} \\
- & \ce{Cm} & bcc & cubic & Cm\_bcc\_eqvol.cif & 229 & This work \\
- & \ce{Cm} & dhcp & hexagonal & Cm\_dhcp\_exp.cif & 194 & \cite{STEVENSON1979201} \\
- & \ce{Cm} & fcc & cubic & Cm\_fcc\_exp.cif & 225 &  \cite{STEVENSON1979201}  \\
- & \ce{Cm2O3} & A-type & trigonal & Cm2O3\_Atype\_P-3m1.cif & 164 & \cite{HAUG19672753} \\
1528054 & \ce{CmO} & NaCl & cubic & CmO\_Fm-3m\_COD1528054.cif & 225 & \cite{Preparationofraremetallicactinides248Cm249Bk249Cfandinvestigationoftheircrystalstructure} \\
9009009 & \ce{CmO2} & fluorite & cubic & CmO2\_Fm-3m\_COD9009009.cif & 225 &  \cite{wyckoff1963fluorite} \\
- & \ce{Fr} & bcc & cubic & Fr\_bcc\_GGA\_primitive\_a6.1924.cif & 229 & \cite{koufos_electronic_2013} \\
- & \ce{Fr} & fcc & cubic & Fr\_fcc\_GGA\_primitive\_a7.8398.cif & 225 & \cite{koufos_electronic_2013} \\
- & \ce{Fr} & hcp & hexagonal & Fr\_hcp\_GGA\_primitive\_a5.5516\_c9.0659.cif & 194 & \cite{koufos_electronic_2013} \\
1509138 & \ce{Po} & alpha & cubic & Po\_alpha\_Pm-3m\_a3.345\_COD1509138\_exp.cif & 221 & \cite{10.1063/1.1747155} \\
- & \ce{Po} & bcc & cubic & Po\_bcc\_Im-3m\_a4.214\_theory-eqvol.cif & 229 & This work \\
1537759 & \ce{Po} & beta & trigonal & Po\_beta\_R-3m\_hex\_COD1537759\_exp.cif & 166 & \cite{10.1063/1.1747155}\\
- & \ce{Po} & fcc & cubic & Po\_fcc\_Fm-3m\_a5.310\_theory-eqvol.cif & 225 & This work \\
- & \ce{Po} & hcp & hexagonal & Po\_hcp\_P63mmc\_a3.755\_c6.131\_theory-eqvol.cif & 194 & This work \\
1010519 & \ce{Po} & monoclinic & monoclinic & Po\_monoclinic\_C121\_COD1010519\_exp.cif & 5 & \cite{10.1063/1.1749762} \\
1534201 & \ce{PoO2} & fluorite & cubic & PoO2\_Fm-3m\_a5.637\_COD1534201\_exp.cif & 225 & \cite{bagnall1954preparation} \\
- & \ce{PoO2} & Pmn21 & orthorhombic & PoO2\_Pmn21\_a3.613\_b5.599\_c4.141.cif & 31 & \cite{MENAD2024e00933} \\
- & \ce{Ra} & bcc & cubic & Ra\_bcc\_Im-3m\_COD1509139\_exp.cif & 229 &  \cite{PhysRevB.82.155121} \\
- & \ce{Ra} & fcc & cubic & Ra\_fcc\_Fm-3m\_a6.5671\_PRB2010\_TableI\_GGA.cif & 225 &  \cite{PhysRevB.82.155121} \\
- & \ce{Ra} & hcp & hexagonal & Ra\_hcp\_P63mmc\_a4.6197\_c7.6202\_PRB2010\_text\_GGA.cif & 194 &  \cite{PhysRevB.82.155121} \\
- & \ce{RaO} & NaCl & cubic & RaO\_Fm-3m\_a5.54\_fallback.cif & 225 & This work  \\
- & \ce{Rn} & bcc & cubic & Rn\_bcc\_est.cif & 229 & This work \\
- & \ce{Rn} & fcc & cubic & Rn\_fcc.cif & 225 & \cite{smits_hundredyearold_2018} \\
- & \ce{Rn} & hcp & Hexagonal & Rn\_hcp.cif & 194 & \cite{smits_hundredyearold_2018} \\
\bottomrule
\end{tabular}
\end{table*}

\vspace{1cm}

\begin{table*}[t]
\centering
\footnotesize
\caption{Literature and database sources for initial Complex Heavy Element (CHE) subset structures.}
\label{tab:4}
\setlength{\tabcolsep}{3pt}
\begin{tabular}{llllHll}
\toprule
COD ID & Primitive Formula & Structure type & Crystal System & & Space Group No. & Ref. \\
\midrule
1010655 & \ce{RaF2} & fluorite & cubic & 1010655/1010655.cif & 225 & \cite{Schulze+1936+430+432} \\
3000125 & \ce{BaRa3S4O16} & barite & orthorhombic & Ra0.76Ba0.24SO4\_Matyskin2017\_Pnma.cif & 62 & \cite{MATYSKIN201715}\\
4519447 & \ce{MgRaC2O6} & - & trigonal & 4519447/4519447.cif & 166 & \cite{doi:10.1021/acsearthspacechem.2c00100}\\
1523087 & \ce{PbPo} & NaCl & cubic & 1523087/1523087.cif & 225 & \cite{witteman1960preparation} \\
1523090 & \ce{HgPo} & NaCl & cubic & 1523090/1523090.cif & 225 & \cite{witteman1960preparation} \\
1525466 & \ce{CdPo} & ZnS & cubic & 1525466/1525466.cif & 216 & \cite{witteman1960preparation}  \\
1530616 & \ce{HfPo} & NiAs & hexagonal & 1530616/1530616.cif & 194 &  \cite{prokin_aksenov_chebotarev_ershova_1978} \\
1534202 & \ce{PoCl2} & - & orthorhombic & 1534202/1534202.cif & 47 & \cite{bagnall1955polonium} \\
1541440 & \ce{Cs2PoBr6} & \ce{K2PtCl6} & cubic & 1541440/1541440.cif & 225 & \cite{bagnall1955poloniumII} \\
4031181 & \ce{LuPo} & NaCl & cubic & 4031181/4031181.cif & 225 & \cite{KERSHNER19661581} \\
4031186 & \ce{HoPo} & NaCl & cubic & 4031186/4031186.cif & 225 &  \cite{KERSHNER19661581}\\
4031200 & \ce{DyPo} & NaCl & cubic & 4031200/4031200.cif & 225 & \cite{KERSHNER19661581}\\
4031201 & \ce{EuPo} & NaCl & cubic & 4031201/4031201.cif & 225 & \cite{KERSHNER19661581}\\
1008976 & \ce{CmAs} & NaCl & cubic & 1008976/1008976.cif & 225 & \cite{osti_4227960} \\
1527406 & \ce{CmBi} & NaCl & cubic & 1527406/1527406.cif & 225 & \cite{GENSINI199375} \\
1535304 & \ce{CmI3O9} & - & monoclinic & 1535304/1535304.cif & 14 & \cite{SYKORA20044413} \\
4118157 & \ce{Cm2B14ClO29} & - & monoclinic & 4118157/4118157.cif & 14 & \cite{doi:10.1021/ja303804r} \\
4346740 & \ce{C28H35Cl4Cm2N4O25} & - & triclinic & 4346740/4346740.cif & 2 & \cite{doi:10.1021/acs.inorgchem.5b02052} \\
9008339 & \ce{CmS2} & \ce{UAs2} & tetragonal & 9008339/9008339.cif & 129 & \cite{oldenbourg1985cu2sb} \\
9008340 & \ce{CmSe2} & \ce{UAs2} & tetragonal & 9008340/9008340.cif & 129 &\cite{oldenbourg1985cu2sb} \\
1552049 & \ce{CfB6H5O13} & - & monoclinic & 1552049/1552049.cif & 15 & \cite{polinski2014unusual} \\
4031557 & \ce{Cf2O2S} & - & trigonal & 4031557/4031557.cif & 164 & \cite{BAYBARZ19742023} \\
4031662 & \ce{CfOF} & fluorite & cubic & 4031662/4031662.cif & 225 & \cite{PETERSON19682955} \\
4300618 & \ce{CfI3O9} & - & monoclinic & 4300618/4300618.cif & 14 & \cite{doi:10.1021/ic051667v} \\
1527345 & \ce{RbAmCO5} & - & hexagonal & 1527345/1527345.cif & 194 & \cite{ellinger1954crystal} \\
1530991 & \ce{AmOBr} & PbFCl & tetragonal & 1530991/1530991.cif & 129 & \cite{WEIGEL197981} \\
1539269 & \ce{Am_2Mo_2O_8} & - & tetragonal & 1539269/1539269.cif & 88 &  \cite{TABUTEAU1978153} \\
1551090 & \ce{AmHWO5} & - & monoclinic & 1551090/1551090.cif & 14 &  \cite{C9SC01174A} \\
1564665 & \ce{C54H78Am2N12O6} & - & monoclinic & 1564665/1564665.cif & 14 &  \cite{D1SC03905A} \\
2018120 & \ce{C6H9AmNaO8} & - & cubic & 2018120/2018120.cif & 198 & \cite{Grigoriev:yf3001} \\
2018121 & \ce{C13H21AmN3O8} & - & monoclinic & 2018121/2018121.cif & 14 & \cite{Grigoriev:yf3001} \\
4031505 & \ce{AmI2} & - & monoclinic & 4031505/4031505.cif & 14 & \cite{BAYBARZ19723427} \\
4031506 & \ce{AmOI} & PbFCl & tetragonal & 4031506/4031506.cif & 129 & \cite{BAYBARZ19723427} \\
4118156 & \ce{AmB9O18} & - & monoclinic & 4118156/4118156.cif & 14 & \cite{doi:10.1021/ja303804r} \\
4124620 & \ce{KAmO2F2} & - & trigonal & 4124620/4124620.cif & 166 &  \cite{Asprey1954PreparationIA} \\
4124687 & \ce{AmOCl} & PbFCl & tetragonal & 4124687/4124687.cif & 129 & \cite{templeton1953crystal} \\
4127110 & \ce{CsAmCr2O8} & - & triclinic & 4127110/4127110.cif & 2 &  \cite{doi:10.1021/jacs.7b09474}\\
4127111 & \ce{CsAmCr2O8} & - & monoclinic & 4127111/4127111.cif & 13 & \cite{doi:10.1021/jacs.7b09474}\\
4300606 & \ce{AmI3O9} & - & monoclinic & 4300606/4300606.cif & 14 & \cite{doi:10.1021/ic0516414} \\
4309330 & \ce{AmI3O9} & - & monoclinic & 4309330/4309330.cif & 14 & \cite{doi:10.1021/ic050386k} \\
4329911 & \ce{Am2P3O10} & - & monoclinic & 4329911/4329911.cif & 15 & \cite{CrossJustinN2012Ssas} \\
4343599 & \ce{Am2H16S3O20} & - & monoclinic & 4343599/4343599.cif & 15 & \cite{doi:10.1021/ic50115a050} \\
4346616 & \ce{C6H12Am2O23} & - & monoclinic & 4346616/4346616.cif & 14 & \cite{doi:10.1021/acs.inorgchem.5b01781} \\
7047397 & \ce{C27H38AmN5S6} & - & monoclinic & 7047397/7047397.cif & 14 & \cite{C8DT02658K} \\
7103819 & \ce{K3Am3HI13O39} & - & trigonal & 7103819/7103819.cif & 161 & \cite{B304530G} \\
7125922 & \ce{C39AmN11S7} & - & cubic & 7125922/7125922.cif & 221 & \cite{C9CC07612C} \\
7130789 & \ce{C15H20AmN3O7S3} & - & triclinic & 7130789/7130789.cif & 2 & \cite{D2CC03352F} \\
7713802 & \ce{C27H27AmN6O9} & - & Monoclinic & 7713802/7713802.cif & 14 & \cite{D3DT02319B} \\
\bottomrule
\end{tabular}
\end{table*}

\FloatBarrier

\footnotesize
\setlength{\LTcapwidth}{\textwidth}
\begin{longtable*}{l l l l H l l}
\caption{\label{tab:5}Literature and additional database sources beyond COD for initial Complex Heavy Element (CHE) subset structures.}\\
\toprule
COD ID & Primitive Formula & Structure type & Crystal System &  & Space Group No. & Ref. \\
\midrule
\endfirsthead

\toprule
COD ID & Primitive Formula & Structure type & Crystal System &  & Space Group No. & Ref. \\
\midrule
\endhead

\midrule
\multicolumn{7}{r}{\emph{Continued on next page}}\\
\endfoot

\bottomrule
\endlastfoot

- & \ce{AmPd3} & \ce{Cu3Au} & cubic & Am0.92Pd3.08\_Cu3Au.cif & 221 &  \cite{VillarsCenzualGladyshevskii+2017} \\
- & \ce{Am12S16} & \ce{Th3P4} & cubic & Am2.67S4\_Th3P4.cif & 220 & \cite{VillarsCenzualGladyshevskii+2017} \\
- & \ce{Am12Se16} & \ce{Th3P4} & cubic & Am3Se4\_Th3P4.cif & 220 & \cite{VillarsCenzualGladyshevskii+2017} \\
- & \ce{Am12Te16} & \ce{Th3P4} & cubic & Am3Te4\_Th3P4.cif & 220 & \cite{VillarsCenzualGladyshevskii+2017} \\
- & \ce{Am12Sb16} & \ce{Th3P4} & cubic & Am4Sb3\_Th3P4.cif & 220 & \cite{VillarsCenzualGladyshevskii+2017} \\
- & \ce{Am4As4} & NaCl & cubic & AmAs\_NaCl.cif & 225 & \cite{VillarsCenzualGladyshevskii+2017} \\
- & \ce{Am4Bi4} & NaCl & cubic & AmBi\_NaCl\_rt.cif & 225 & \cite{VillarsCenzualGladyshevskii+2017} \\
- & \ce{Am8Fe16} & \ce{MgCu2} & cubic & AmFe2\_MgCu2.cif & 227 & \cite{VillarsCenzualGladyshevskii+2017} \\
- & \ce{Am4H8} & \ce{CaF2} & cubic & AmH2\_CaF2.cif & 225 & \cite{VillarsCenzualGladyshevskii+2017} \\
- & \ce{Am8Ir16} & \ce{MgCu2} & cubic & AmIr2\_MgCu2.cif & 227 & \cite{VillarsCenzualGladyshevskii+2017} \\
- & \ce{AmLa3} & \ce{Cu3Au} & cubic & AmLa3\_Cu3Au.cif & 221 & \cite{VillarsCenzualGladyshevskii+2017} \\
- & \ce{Am12N16} & \ce{Th3P4} & cubic & AmN1.25\_Th3P4.cif & 220 & \cite{VillarsCenzualGladyshevskii+2017} \\
- & \ce{Am4N4} & NaCl & cubic & AmN\_NaCl.cif & 225 & \cite{VillarsCenzualGladyshevskii+2017} \\
- & \ce{Am8Ni16} & \ce{MgCu2} & cubic & AmNi2\_MgCu2.cif & 227 & \cite{VillarsCenzualGladyshevskii+2017} \\
- & \ce{Am4O4} & NaCl & cubic & AmO\_NaCl.cif & 225 & \cite{VillarsCenzualGladyshevskii+2017} \\
- & \ce{Am8Rh16} & \ce{MgCu2} & cubic & AmRh2\_MgCu2.cif & 227 & \cite{VillarsCenzualGladyshevskii+2017} \\
- & \ce{AmRh3} & \ce{Cu3Au} & cubic & AmRh3\_Cu3Au.cif & 221 & \cite{VillarsCenzualGladyshevskii+2017} \\
- & \ce{Am4S4} & NaCl & cubic & AmS\_NaCl.cif & 225 & \cite{VillarsCenzualGladyshevskii+2017} \\
- & \ce{AmSi2} & \ce{AlB2} & hexagonal & AmSi1.67\_AlB2.cif & 191 & \cite{VillarsCenzualGladyshevskii+2017} \\
- & \ce{AmTe} & CsCl & cubic & AmTe\_CsCl\_hp.cif & 221 & \cite{VillarsCenzualGladyshevskii+2017} \\
- & \ce{Am4Te4} & NaCl & cubic & AmTe\_NaCl.cif & 225 & \cite{VillarsCenzualGladyshevskii+2017} \\
- & \ce{AmCu5} & \ce{CaCu5} & hexagonal & Cu5Am\_CaCu5.cif & 191 & \cite{VillarsCenzualGladyshevskii+2017} \\
- & \ce{AmB4} & \ce{UB4} & tetragonal & AmB4\_UB4\_P4mbm\_Zalkin1953.cif & 127 & \cite{Zalkin:a00862} \\
- & \ce{AmB6} & \ce{CaB6} & cubic & AmB6\_CaB6\_cP7\_prototype.cif & 221 & \cite{VillarsCenzualGladyshevskii+2017} \\
- & \ce{AmCl} & CsCl & cubic & AmCl\_CsCl\_B2\_prototype.cif & 221 & \cite{VillarsCenzualGladyshevskii+2017} \\
- & \ce{AmSb2} & \ce{SmSb2} & orthorhombic & AmSb2\_SmSb2\_Cmca\_Wang1967.cif & 64 & \cite{wang1967crystal} \\
- & \ce{Cf4S4} & NaCl & cubic & CfS\_NaCl\_Handbook2017.cif & 225 & \cite{VillarsCenzualGladyshevskii+2017} \\
- & \ce{Cf4Sb4} & NaCl & cubic & CfSb\_NaCl\_Handbook2017.cif & 225 & \cite{VillarsCenzualGladyshevskii+2017} \\
- & \ce{CmCo2} & \ce{MgCu2} & cubic & CmCo2\_MgCu2.cif & 227 & \cite{VillarsCenzualGladyshevskii+2017} \\
- & \ce{CmCo5} & \ce{CaCu5} & hexagonal & CmCo5\_CaCu5.cif & 191 & \cite{VillarsCenzualGladyshevskii+2017} \\
- & \ce{CmFe2} & \ce{MgCu2} & cubic & CmFe2\_MgCu2.cif & 227 & \cite{VillarsCenzualGladyshevskii+2017} \\
- & \ce{CmH2} & \ce{CaF2} & cubic & CmH2\_CaF2.cif & 225 & \cite{VillarsCenzualGladyshevskii+2017} \\
- & \ce{CmIr3} & \ce{Cu3Au} & cubic & CmIr3\_Cu3Au.cif & 221 & \cite{VillarsCenzualGladyshevskii+2017} \\
- & \ce{CmN} & NaCl & cubic & CmN\_NaCl.cif & 225 &\cite{VillarsCenzualGladyshevskii+2017} \\
- & \ce{CmNi3} & \ce{Cu3Au} & cubic & CmNi3\_Cu3Au.cif & 221 & \cite{VillarsCenzualGladyshevskii+2017} \\
- & \ce{CmNi5} & \ce{CaCu5} & hexagonal & CmNi5\_CaCu5.cif & 191 & \cite{VillarsCenzualGladyshevskii+2017} \\
- & \ce{CmP} & NaCl & cubic & CmP\_NaCl.cif & 225 & \cite{VillarsCenzualGladyshevskii+2017} \\
- & \ce{CmPt3} & \ce{Cu3Au} & cubic & CmPt3\_Cu3Au.cif & 221 & \cite{VillarsCenzualGladyshevskii+2017} \\
- & \ce{CmPt5} & \ce{CaCu5} & hexagonal & CmPt5\_CaCu5\_hex.cif & 191 &\cite{VillarsCenzualGladyshevskii+2017} \\
- & \ce{CmRh2} & \ce{MgCu2} & cubic & CmRh2\_MgCu2.cif & 227 & \cite{VillarsCenzualGladyshevskii+2017} \\
- & \ce{CmRh3} & \ce{Cu3Au} & cubic & CmRh3\_Cu3Au.cif & 221 & \cite{VillarsCenzualGladyshevskii+2017} \\
- & \ce{CmS} & NaCl & cubic & CmS\_NaCl.cif & 225 & \cite{VillarsCenzualGladyshevskii+2017} \\
- & \ce{CmSb} & NaCl & cubic & CmSb\_NaCl.cif & 225 &\cite{VillarsCenzualGladyshevskii+2017} \\
- & \ce{CmSe} & NaCl & cubic & CmSe\_NaCl.cif & 225 & \cite{VillarsCenzualGladyshevskii+2017} \\
- & \ce{CmTe} & NaCl & cubic & CmTe\_NaCl.cif & 225 & \cite{VillarsCenzualGladyshevskii+2017} \\
- & \ce{CmH3} & \ce{Na3As} & hexagonal & CmH3\_Na3As\_D018\_P63mmc.cif & 194 & \cite{VillarsCenzualGladyshevskii+2017} \\
- & \ce{CmRu2} & \ce{MgZn2} & hexagonal & CmRu2\_MgZn2\_C14\_P63mmc.cif & 194 & \cite{VillarsCenzualGladyshevskii+2017} \\
- & \ce{FrBeBr3} & perovskite & cubic & FrBeBr3\_Pm-3m\_0GPa.cif & 221 & \cite{https://doi.org/10.1002/nano.70095} \\
- & \ce{FrBeCl3} & perovskite & cubic & FrBeCl3\_Pm-3m\_0GPa.cif & 221 & \cite{https://doi.org/10.1002/nano.70095} \\
- & \ce{FrCaBr3} & perovskite & cubic & FrCaBr3\_Pm-3m\_0GPa\_PBE.cif & 221 & \cite{mansur2025pressure} \\
- & \ce{FrCaCl3} & perovskite & cubic & FrCaCl3\_Pm-3m\_0GPa\_PBE.cif & 221 & \cite{AHMED2024e34059, mansur2025pressure}  \\
- & \ce{FrCaI3} & perovskite & cubic & FrCaI3\_Pm-3m\_0GPa\_PBE.cif & 221 & \cite{mansur2025pressure} \\
- & \ce{FrSrBr3} & perovskite & cubic & FrSrBr3\_Pm-3m\_0GPa.cif & 221 &\cite{10.1063/5.0281327} \\
- & \ce{FrSrCl3} & perovskite & cubic & FrSrCl3\_Pm-3m\_0GPa.cif & 221&\cite{10.1063/5.0281327} \\
- & \ce{FrSrF3} & perovskite & cubic & FrSrF3\_Pm-3m\_0GPa.cif & 221 &\cite{10.1063/5.0281327} \\
- & \ce{FrSrI3} & perovskite & cubic & FrSrI3\_Pm-3m\_0GPa.cif & 221 &\cite{10.1063/5.0281327} \\
- & \ce{FrZnBr3} & perovskite & cubic & FrZnBr3\_Pm-3m\_0GPa\_PBE.cif & 221 & \cite{https://doi.org/10.1155/er/5549809} \\
- & \ce{FrZnCl3} & perovskite & cubic & FrZnCl3\_Pm-3m\_0GPa\_PBE.cif & 221 & \cite{https://doi.org/10.1155/er/5549809} \\
- & \ce{FrZnF3} & perovskite & cubic & FrZnF3\_Pm-3m\_0GPa\_PBE.cif & 221 & \cite{https://doi.org/10.1155/er/5549809} \\
- & \ce{FrGeBr3} & perovskite & cubic & FrGeBr3\_Pm-3m\_0GPa.cif & 221 &\cite{10.1063/5.0201448} \\
- & \ce{FrGeCl3} & perovskite & cubic & FrGeCl3\_Pm-3m\_0GPa.cif & 221 & \cite{10.1063/5.0201448} \\
- & \ce{FrGeI3} & perovskite & cubic & FrGeI3\_Pm-3m\_0GPa.cif & 221 & \cite{10.1063/5.0201448} \\
- & \ce{FrSnBr3} & perovskite & cubic & FrSnBr3\_Pm-3m\_0GPa.cif & 221 & \cite{10.1063/5.0207336} \\
- & \ce{FrSnCl3} & perovskite & cubic & FrSnCl3\_Pm-3m\_0GPa.cif & 221 & \cite{10.1063/5.0207336} \\
- & \ce{FrSnI3} & perovskite & cubic & FrSnI3\_Pm-3m\_0GPa.cif & 221 & \cite{10.1063/5.0207336} \\
- & \ce{RaS} & NaCl & cubic & RaS\_Handbook2017\_NaCl\_Fm-3m.cif & 225 & \cite{VillarsCenzualGladyshevskii+2017} \\ 
- & \ce{RaSe} & NaCl & cubic & RaSe\_Handbook2017\_NaCl\_Fm-3m.cif & 225 & \cite{VillarsCenzualGladyshevskii+2017} \\ 
- & \ce{RaCrO4} & barite & orthorhombic & RaCrO4\_Butkalyuk2021\_Pnma\_MODEL.cif & 62 & \cite{butkalyuk2021study} \\
- & \ce{RaPbO3} & perovskite & cubic & RaPbO3\_Butkalyuk2013\_Pm3m.cif & 221 & \cite{butkalyuk2013synthesis} \\
- & \ce{RaSO4} & barite & orthorhombic & RaSO4\_Matyskin2017\_Pnma.cif & 62 & \cite{MATYSKIN201715} \\
- & \ce{RaWO4} & scheelite & tetragonal & RaWO4\_RomeroVazquez2023\_I41a\_DFT.cif & 88 &\cite{ROMEROVAZQUEZ2023123709} \\
- & \ce{RnF2} & \ce{XeF2} & tetragonal & RnF2\_Liao1998\_Bondi.cif & 139 & \cite{doi:10.1021/jp9825516} \\
\end{longtable*}

\endgroup
\end{widetext}
}

\end{document}